%% file: RD-revised.tex
\documentclass[10pt,journal,cspaper,compsoc]{IEEEtran}
\usepackage{graphicx}
\usepackage{epstopdf}
\usepackage{amsmath,amsthm}
\usepackage{url}
\usepackage{caption}
\usepackage{subcaption}
\usepackage{tabularx}
\usepackage{multirow}
\usepackage{booktabs}
\newcommand{\bi}{\begin{itemize}}
\newcommand{\ei}{\end{itemize}}
\newcommand{\be}{\begin{enumerate}}
\newcommand{\ee}{\end{enumerate}}

\newtheorem{lemma}{Lemma}
\newtheorem{theorem}{Theorem}
\newtheorem{corollary}{Corollary}

\DeclareMathOperator*{\argmax}{arg\,max}

\begin{document}
%
\title{Max-Sum Diversification, Monotone Submodular Functions and Dynamic Updates}
%
%
%
%

\author{Allan Borodin
        and Aadhar Jain and Hyun Chul Lee and Yuli Ye
\IEEEcompsocitemizethanks{\IEEEcompsocthanksitem A. Borodin is with the Department
of Computer Science, University of Toronto, Toronto,
Ontario, Canada, M5S 3G4, E-mail: bor@cs.toronto.edu
\IEEEcompsocthanksitem A. Jain and H.C. Lee are with MyFitnessPal, 525 Brannan Street, San Francisco, CA, 94107, E-mails: ajain@myfitnesspal.com , clee@myfitnesspal.com
\IEEEcompsocthanksitem Y. Ye is with Wish, 1 Sansome Street, 40th Floor, San Francisco, CA, 94104
, E-mail: yuli@wish.com
}
\thanks{}}

%
%

\markboth{.}
%
{Shell \MakeLowercase{\textit{et al.}}: Bare Demo of IEEEtran.cls for Computer Society Journals}
%


\IEEEcompsoctitleabstractindextext{%
\begin{abstract}
Result diversification is an important aspect in web-based search, 
document summarization, facility location, portfolio 
management and other applications. 
Given a set of ranked results for a set of objects 
(e.g. web documents, facilities, etc.) with a distance between any pair, 
the goal is to select a subset $S$ satisfying the following three criteria:
(a) the subset $S$   
satisfies some constraint (e.g. bounded cardinality);
(b) the subset contains results of high ``quality''; and
(c) the subset contains results that are ``diverse'' 
relative to the distance measure.
The goal of result diversification is to produce a diversified subset 
while maintaining high quality as much as possible. We study a broad class of problems 
where the distances are a metric,  where the constraint is given by independence in a matroid, where quality is 
determined by a monotone submodular function, and diversity is defined as
the sum of distances between objects in $S$. Our problem is a generalization of the {\em max sum diversification} 
problem studied in \cite{GoSh09} which in turn is a generalization of the
{\em max sum $p$-dispersion problem} studied extensively in location theory. It is NP-hard even with the triangle inequality. We propose two simple and natural algorithms: a greedy algorithm for a cardinality constraint and a local search algorithm for an arbitrary matroid constraint. We prove that both algorithms achieve constant approximation ratios. 
\end{abstract}

}

\maketitle

\IEEEdisplaynotcompsoctitleabstractindextext

%
\IEEEpeerreviewmaketitle

\input{intro}
\input{related}

\input{formu}

\input{submod-rev}
\input{matroid}

\input{dynup}
\input{experiments_tkde}

\input{conclusion-rev}

\bibliographystyle{IEEEtran}
\bibliography{RD-TKDE} 

\appendix
\input{appendix}

\end{document}

%% file: intro.tex
\section{Introduction}

Result diversification has many important applications in databases, operations research, information retrieval, and finance. In this paper, we study and extend a particular version of result diversification, known as max-sum diversification. More specifically, we consider the setting where we 
are given a set of elements in a metric space and a set valuation function $f$ defined on every subset. For any given 
subset $S$, the overall objective is a linear combination of $f(S)$ and 
the sum of the distances induced by $S$. The goal is to find a subset 
$S$ satisfying some constraints that maximizes the overall objective.

This diversification problem is first studied by Gollapudi and Sharma in~\cite{GoSh09} for 
modular (i.e. linear) set functions and for sets satisfying a 
cardinality constraint (i.e. a 
uniform matroid).   (See \cite{GoSh09} for some closely related work.)
The max-sum $p$-dispersion problem seeks to find a subset $S$
of cardinality $p$ so as to maximize $\sum_{x,y \in S} d(x,y)$.
The diversification problem is then a  linear combination of a quality function
$f()$ and the max-sum dispersion function.
Gollapudi and Sharma give a 2 approximation greedy algorithm for
some metrical distance diversification problems   
by reducing to the analogous dispersion problem. More specifically for max-sum
diversification they   
use the greedy algorithm of Hassin, Rubsenstein and Tamir \cite{HassinRT97}.
Hassin et al give a 
non greedy algorithm 
for a more general problem where the goal is to construct 
$k$ subsets each having $p$ elements. (We willl restrict attention to
the case $k = 1$.) Their non greedy algorithm obtains the 
ratio $2-\frac{1}{\lceil p/2 \rceil}$
and hence the same
approximation holds for the Gollapudi and Sharma diversification problem. 

The first part of our paper considers an extension of the modular case to the monotone submodular case,
for which the algorithm in~\cite{GoSh09} no longer applies. 
We are able to maintain the same 2-approximation using 
a natural, but different greedy algorithm. We then further extend the 
problem by considering any matroid constraint and show that a natural 
single swap local search algorithm provides a 2-approximation in this more general setting. This extends the Nemhauser, Wolsey and Fisher~\cite{NWF78} approximation result for the problem of submodular function maximization subject to a matroid constraint (without the distance function component).   
We note that the dispersion function is a supermodular function 
\footnote{Motivated by the analysis in this paper, Borodin et al \cite{BorodinLY14} introduce the
class of {\it weakly submodular functions} and show that the max-sum dispersion
measure as well as all monotone submodular functions are weakly submodular. 
Furthermore, it is shown that the problem of maximizing such functions
subject to cardinality (resp. general matroid) constraints can be 
polynomial time approximated 
within a constant factor by a greedy (resp. local search) algorithm.} 
and hence
the Nemhauser er al result does not immediately extend to our diversification
problem. 

\vspace{.1in}

\vspace{.1in}

Submodular functions have been extensively considered
since they model many natural phenomena. For example, 
in terms of keyword based search in database systems, 
it is well understood that users begin to gradually (or sometimes abruptly) 
lose interest 
the more results they have to consider \cite{vieira11_2, vieira11}. But on the other hand, 
as long as a user continues to gain some benefit,  additional query results can
improve the overall quality but at a decreasing rate. In a related application, 
Lin and Bilnes \cite{LinB2011} argue that monotone submodular functions 
are an ideal class of functions for text summarization.   Following 
and extending the results in \cite{GoSh09},
we consider the case of maximizing a linear combination of
a submodular quality function $f(S)$ and the max-sum dispersion subject to 
a cardinality constraint (i.e., $|S| \leq p$ for some given $p$). 
We present a greedy algorithm that is somewhat unusual in that it does not
try to optimize the objective in each iteration but rather optimizes  a closely
related potential function. We show that our greedy approach  matches 
the greedy $2$-approximation
\footnote{Clearly, in the modular case 
for $p$ constant, a brute force trial of all subsets of
size $p$ is an optimum, albeit
inefficient, algorithm.} 
in \cite{GoSh09} obtained for diversification with a modular quality function.
We note that the greedy algorithm in \cite{GoSh09} utilizes
the max dispersion algorithm of Hassin, Rubinstein and Tamir
\cite{HassinRT97} which greedily adds edges whereas our algorithm
greedily adds vertices. 

Our next result continues with the submodular case
but now we go beyond a cardinality constraint (i.e., the uniform matroid)
 on $S$ and allow the constraint to be that $S$ is independent in 
a given matroid. This allows a substantial increase in generality. 
For example, while diversity might represented by the distance 
between retrieved database tuples under a given criterion 
(for instance, a kernel based diversity measure 
called \textit{answer tree kernel} is used in \cite{fengzhao11}), we could use a partition matroid
to insure that (for example) the retrieved database tuples come from a variety 
of  different sources. That is, we may wish to have 
$n_i$ tuples from a specific database field $i$. This is, of course, another form 
of diversity but one orthogonal to diversity based on the given criterion. Similarly 
in the stock portfolio example, we might wish to have a balance
of stocks in terms of say risk and profit profiles (using some statistical
measure of distances) while using a submodular quality function to reflect 
a users submodular utility for profit and using a partition 
matroid to insure  that different sectors of 
the economy are well represented. Another important class of matroids 
(relevant to the above applications) is that 
of transversal matroids. Suppose we have a collection 
$\{C_1, C_2, \ldots, C_m$\} of (possibly) {\it overlapping} sets (i.e., the collection is
not necessarily a partition) of  
database tuples (or stocks). Our goal might be to derive a set $S$
such that the database tuples in $S$ form a set of representatives for 
the collection; that is, every database tuple in $S$ represents (and is in) a unique
set $C_i$ in the collection.   The set $S$ is then an 
independent set in the
transversal matroid induced by the collection. We also note 
\cite{schrijver03} that
the intersection of any matroid with a uniform matroid is still a matroid
so that in the above examples, we could further impose the constraint   
that the set S has at most $p$ elements. 

Our final theoretical result concerns dynamic updates. Here we 
restrict attention to a modular set function $f(S)$; that is, we now
have weights on the elements and $f(S) = \sum_{u \in S} w(u)$ 
where $w(u)$ is the weight of element $u$. This allows us to consider 
changes to the weight of a single element as well as changes to the 
distance function. 

The rest of the paper is organized as follows. 
In Section 2, we discuss related work in 
dispersion and result diversification. 
In Section 3, we formulate the problem as a combinatorial optimization problem
and discuss the complexity of the problem.
In Section 4, we consider max-sum diversification with monotone submodular 
set quality functions subject to a cardinality constriant and give a 
conceptually simple greedy algorithm that achieves a 2-approximation.
We extend the problem to the matroid case in Section 5 and discuss dynamic 
updates in Section 6. Section 7 carries out a number of experiments. In
particular, we compare our greedy algorithm with the greedy 
algorithm of Gollapudi and Sharma. Section 8 concludes the paper.

\vspace{.1in} 

%% file: related.tex
\section{Related Work}
With the proliferation of today's social media, database and web content, ranking becomes an important problem as it decides what gets selected and what does not, and what to be displayed first and what to be displayed last.
Many early ranking algorithms, for example in web search, are based on the notion of ``relevance", i.e., the closeness of the object to the search query.  
However, there has been a rising interest to incorporate some notion of ``diversity" into 
measures of quality.

One early work in this direction is the notion of ``Maximal Marginal Relevance'' (MMR)  introduced by Carbonell
and Goldstein in~\cite{Carbonell:1998:UMD:290941.291025}. More specifically, MMR is defined as follows:
$${\rm MMR}=\max_{D_i\in R\setminus S}[\lambda\cdot sim_1(D_i, Q)-(1-\lambda) \max_{D_j\in S}sim_2(D_i,D_j)],$$
where $Q$ is a query; $R$ is the ranked list of documents retrieved; $S$ is the subset of documents in
$R$ already selected; $sim_1$ is the similarity measure between a document and a query, and
$sim_2$ is the similarity measure between two documents. The parameter $\lambda$ controls the trade-off
between novelty (a notion of diversity) and relevance. The MMR algorithm iteratively selects the next document with 
respect to the MMR objective function until a given cardinality condition is met. 
The MMR heuristic has been widely used, but to the best of our knowledge, it has not been theoretically justified. 
Our paper provides some theoretical evidence why MMR is a legitimate approach for diversification.
The greedy algorithm we propose in this paper can be viewed as a natural extension of MMR.

There is extensive research on how to diversify returned ranking results to satisfy multiple users. Namely, the result diversity issue occurs when many facets of queries are discovered and a set of multiple users expect to find their desired facets in the first page of the results. Thus, the challenge is to find the best strategy for ordering the results such that many users would find their relevant pages in the top few slots. 

Rafiei et al.~\cite{DBLP:conf/www/RafieiBS10} modeled this as a continuous optimization problem. They introduce a weight vector $W$ for the search results, where the total weight sums to one. 
They define the portfolio variance to be $W^TCW$, where $C$ is the covariance matrix of the result set. The goal then is to minimize the portfolio variance while the expected relevance is fixed at a certain level. They report that their proposed algorithm can improve upon Google in terms of the diversity on random queries, retrieving $14\%$ to $38\%$ more aspects of queries in  top five, while maintaining a precision very close to Google. 

Bansal et al.~\cite{DBLP:conf/icalp/BansalJKN10} considered the setting in which various types of users exist and each is interested in a subset of the search results. They use a performance measure based on {\em discounted cumulative gain}, which defines the usefulness (gain) of a document as its position in the resulting list. 
Based on this measure, they suggest a general approach to develop approximation algorithms for ranking search results that captures different aspects of users' intents.
They also take into account that the relevance of one document cannot be treated independent of the relevance of other documents in a collection returned by a search engine.
They consider both the scenario where users are interested in only a single search result (e.g., navigational queries)
and the scenario where users have different requirements on the number of search results, and develop good approximation solutions for them. 

The database community has recently studied the query diversification problem, which is mainly for keyword 
search in databases \cite{liu09, Yu:2009,drosou09,vieira11, fengzhao11,vieira11_2, Demidova10}.
Given a very large database, an exploratory query can easily lead to a vast answer set. Typically, an answer's relevance to the user query is based on \textit{top-k} or \textit{tf-idf}. As a way of increasing user 
satisfaction, different query diversification techniques have been proposed including some system based ones taking into account query parameters, evaluation algorithms, 
and dataset properties. For many of these, a max-sum type objective function is usually used. 

Other than those discussed above, there are many recent papers studying result diversification in different settings, via different approaches and through different perspectives, for example~\cite{DBLP:conf/sigir/ZhaiCL03,DBLP:conf/sigir/ChenK06,DBLP:conf/naacl/ZhuGGA07,DBLP:conf/icml/YueJ08,DBLP:conf/icml/RadlinskiKJ08,DBLP:conf/wsdm/AgrawalGHI09,DBLP:conf/wsdm/BrandtJYB11,SantosMO11,DBLP:conf/wsdm/DouHCSW11,DBLP:conf/icml/SlivkinsRG10}. 
The reader is referred to~\cite{DBLP:conf/wsdm/AgrawalGHI09,DrosouP10} for a good summary of the field. 
Most relevant to our work is the paper by Gollapudi and Sharma~\cite{GoSh09}, where they develop an axiomatic approach to characterize and design diversification systems. Furthermore, they consider three different 
diversification objectives and using earlier results in facility dispersion, they are able to give algorithms with good worst case approximation guarantees. 
This paper is a continuation of research along this line.

Recently, Minack et al.~\cite{minack11} have studied the problem of incremental diversification for very large data sets.
Instead of viewing the input of the problem as a set, they consider the input as a stream, and use a simple online algorithm to process each element in an incremental fashion, maintaining a near-optimal diverse set at any point in the stream. Although their results are largely experimental, this approach significantly reduces CPU and memory consumption, and hence is applicable to large data sets. Our dynamic update algorithm deals with a problem of a similar nature, but in addition to our experimental results, we are also able to prove theoretical guarantees. To the best of our knowledge, 
our work is the first of its kind to obtain a near-optimality condition for result diversification in a dynamically changing environment.

Independent of our conference paper \cite{BorodinLY12}, Abbassi, Mirrokni and
Thakus \cite{AbbassiMT13} have also shown that the (Hamming distance 1) local search algorithm provides a 2-approximation
for the max-sum dispersion problem subject to a matroid constraint. Their
version of the dispersion problem is somwehat more general in that
they additionally consider that the points are chosen from different
clusters. They indirectly consider a quality measure
by first restricting the universe of objects to high quality objects
and then apply dispersion. They provide a number of interesting experimental
results.



%% file: formu.tex
\section{Problem Formulation}
\label{sec:formu}
Although the notion of ``diversity" naturally arises in the context of databases, social media and web search, 
the underlying mathematical object is not new.
As presented in~\cite{GoSh09}, there is a rich and long line of research 
in location theory dealing with a similar concept; in particular,
one objective is the placement of facilities on a network to maximize some function of the distances between facilities. The situation arises when 
proximity of facilities is undesirable, for example, the distribution of business franchises in a city. Such location problems are often referred to as {\em dispersion} problems; for more 
motivation and early work, see~\cite{RePEc:eee:ejores:v:46:y:1990:i:1:p:48-60, RePEc:eee:ejores:v:40:y:1989:i:3:p:275-291, GEAN:GEAN133}.

Analytical models for the dispersion problem assume that the given network is represented by a set 
$V=\{v_1,v_2, \dots, v_n\}$ of $n$ vertices along with a  distance function
between every pair of vertices. The objective is to locate 
$p$ facilities ($p\le n$)
among the $n$ vertices, with at most one facility per vertex, such that some function of distances between facilities is maximized.  Different objective functions are considered for the dispersion problems in the literature including: the  max-sum criterion (maximize the total distances between all pairs of facilities) in~\cite{Wang:1988:STG:49310.49312,RePEc:eee:ejores:v:46:y:1990:i:1:p:48-60,RRT94}, the max-min criterion (maximize the minimum distance between a pair of facilities) in~\cite{GEAN:GEAN133,RePEc:eee:ejores:v:46:y:1990:i:1:p:48-60,RRT94}, the max-mst (maximize the minimum spanning tree among all facilities) and many other related criteria in~\cite{HIKT95, DBLP:journals/jal/ChandraH01}. 
When the distances are arbitrary, the max-sum 
problem is a weighted generaliztion 
of the densest subgraph problem which is a known difficult problem 
not admitting a PTAS~(\cite{Khot06} and not known to have a constant approximation algorithm. Sometimes the problem is studied for specific metric distances 
(e.g as in Fekete and Meijer \cite{FeketeM03}) or 
for restricted classes of weights (e.g. as in Czygrinow \cite{Czygrinow00}) 
where there can be a PTAS.    

Our diversification problem is a generalization of the 
max sum $p$-dispersion problem assuming 
arbitrary metric distances. For the max-sum criteria and for most of the 
objective criteria, the dispersion problem 
is NP-hard, and approximation algorithms have been developed and studied; see~\cite{DBLP:journals/jal/ChandraH01} for a summary of known results. 
Our diversification problem is a generalization of the following 
max sum $p$-dispersion problem for 
arbitrary metric distances. Most relevant to this paper is the max-sum dispersion problem with metric distances.

{\sc Problem 1.} {\tt Max-Sum $p$ Dispersion} \\
\\
Let $U$ be the underlying ground set, and let $d(\cdot,\cdot)$ be a metric
distance function on $U$.
Given a fixed integer $p$, the goal of the problem is to find a subset $S \subseteq U$
that:
$$\begin{array}{ll}
\rm{maximizes} & \sum_{\{u,v\}:{u,v\in S}} d(u,v)\\
\\
\rm{subject\;\; to} & |S|=p,
\end{array}$$

The problem is known to be NP-hard
by an easy reduction from Max-Clique, and as noted by 
Alon \cite{Alon14}, there is evidence that the problem is hard to 
compute in polynomial time with approximation
$2-\epsilon$ for any $\epsilon > 0$
when $p = n^r$ for $1/3 \leq r < 1$.  
Namely, based on the assumption that the planted clique problem is hard, 
Alon et al \cite{Alonammw11}
show that it is hard to distinguish between a graph having a large 
planted clique of
size $p$ 
and one in which the densest subgraph of size $p$ is of 
density at most an arbitrarily small constant 
$\delta$ (for suffiently large $n$). 
Considering the complement of a random graph $G$ in ${\cal G}(n,1/2)$,
their result says that it
is hard to distinguish between a graph having an independent set of size $p$ and
one in which the density of edges in any size $p$-subgraph is at least 
$(1-\delta)$.
Adding another node to the complement graph that is connected
to all nodes in $G$,  
the graph distance metric is now the $\{1,2\}$ metric formed by the 
transitive closure so that
adjacent nodes have distance 1 and non adjacent nodes have distance 2.
So we therefore cannot distinguish between graphs where there exists a set of 
nodes $S$ of size $p$ ( for
$p$ as above) where $\sum_{(u,v) \in S} d(u,v) = {p \choose 2} * 2$ and one
where in every set of size $p$, we have
$\sum_{(u,v) \in S} d(u,v) \leq {p \choose 2} [(1-\delta) + 2 \delta]$.

In~\cite{RRT94}, Ravi, Rosenkrantz and Tayi give a greedy algorithm (greedily 
choosing vertices 
that is shown to have approximation ratio no worse than $4$ and no better 
than $\frac{2}{1+2/p(p-1)}$. 
Hassin, Rubenstein and Tamir \cite{HassinRT97} improve upon the Ravi et al 
result by 
an algorithm that greedily chooses {\it edges} 
yielding an approximation ratio of $2$. 
Hassin et al also give an algorithm based
on maximum matching that provides a $2-\frac{1}{\lceil p/2 \rceil}$ 
approximation for a more
general problem; 
namely, the algorithm must find a subset $U'$ which 
is partitioned 
into $k$ disjoint subsets, each of
size $p$ so as to maximize the pairwise sum of all pairs of vertices in $U'$. 
The more general $(p,k)$ problem is similar to a
partition matroid constraint but in a partition matroid, the partition
is given as part of the definition of the matroid and each block
of the partition has its own cardinality constraint.

Answering an open problem stated
in Hassin et al., Birnbaum and Goldman \cite{BirnbaumG09} give an 
improved analysis proving that 
the Ravi et al greedy algorithm results in a $\frac{2p-2}{p-1}$ approximation 
for the max-sum $p$ dispersion problem. This then shows that a 2-approximation 
is a tight bound (as $p$ grows) for the Ravi et al greedy algorithm. 
More generally, Birnbaum and Goldman show that greedily choosing a
set of $d$ nodes provides a $\frac{2p-2}{p+d-2}$ approximation. Our
analysis in Section~\ref{sec:submo} yields an alternative proof 
that the Ravi et al greedy
algorithm approximation ratio is no worse than $2$ even when extended to
the max-sum $p$ diversification problem (with a monotone submodular 
value function) considered in Section~\ref{sec:submo}.

\medskip 

{\sc Problem 2.} {\tt Max-Sum $p$ Diversification} \\ 
\\
Let $U$ be the underlying ground set, and let $d(\cdot,\cdot)$ be a metric 
distance function on $U$.
For any subset of $U$, let $f(\cdot)$ be 
a non-negative set function measuring 
the value of a 
subset. Given a fixed integer $p$, the goal of the problem is to find a subset $S \subseteq U$ 
that:
$$\begin{array}{ll}
\rm{maximizes} & f(S)+\lambda\sum_{\{u,v\}:{u,v\in S}} d(u,v)\\
\\
\rm{subject\;\; to} & |S|=p,
\end{array}$$
where $\lambda$ is a parameter specifying a desired trade-off between the two objectives.

The max-sum diversification problem is first proposed and studied in the context of result diversification in~\cite{GoSh09}~\footnote{In fact, they have a slightly different but equivalent formulation.}, where the function $f(\cdot)$ is modular.  
In their paper, the value of $f(S)$ measures the relevance of a given subset to a search query, and the value $\sum_{\{u,v\}:{u,v\in S}} d(u,v)$ gives a diversity measure on $S$. The parameter $\lambda$ specifies a desired trade-off between diversity and relevance. They reduce the problem to the max-sum dispersion problem, and using an algorithm in~\cite{HassinRT97}, they obtain an approximation ratio of 2.

In this paper, we first study the problem with more general valuation functions;namely, normalized, monotone submodular set functions. 
For notational convenience, for any two sets $S$, $T$ and an element $e$, we write $S\cup\{e\}$ as $S+e$, $S\setminus\{e\}$ as $S-e$, $S\cup T$ as $S+T$, 
and $S\setminus T$ as $S-T$.
A set function $f$ is {\em normalized} if $f(\emptyset)=0$.
The function is {\em monotone} if for any $S, T\subseteq U$ and $S\subseteq T$, $$f(S)\le f(T).$$ It is {\em submodular} if for any $S, T\subseteq U$, $S\subseteq T$ with $u\in U$, $$f(T+u)-f(T)\le f(S+u)-f(S).$$
In the remainder of paper, all functions considered are normalized. 

We proceed to our 
first contribution, a greedy algorithm (different than the one 
in~\cite{GoSh09}) that obtains a 2-approximation for 
monotone submodular 
set functions. 



%% file: submod-rev.tex
\section{Submodular Functions}
\label{sec:submo}
Submodular set functions can be characterized by
the property of a decreasing marginal gain as the size of the set increases.
%
As such, submodular functions are well-studied objects in economics, game theory and combinatorial optimization.
More recently, submodular functions have attracted attention in many practical fields of computer science.
For example, Kempe et al.~\cite{Kempe:2003:MSI:956750.956769} study the problem of selecting a set of most influential nodes to maximize
the total information spread in a social network. They have shown that under two basic stochastic diffusion models, the expected influence of an initially chosen 
set is submodular,
hence the problem admits a good approximation algorithm. 
In natural language processing, Lin and Bilmes \cite{LinBX2009,LinB10,LinB2011} have studied a class of submodular functions for document summarization. 
These functions each combine two terms, one which encourages the summary to be representative of the corpus, and the other which positively rewards diversity. 
Their experimental results show that a greedy algorithm with the objective of maximizing these submodular functions outperforms the existing state-of-art results
in both generic and query-focused document summarization.  

Both of the above mentioned results are based on the fundamental work of
Nemhauser, Wolsey and Fisher~\cite{NWF78}, which gave 
an $\frac{e}{e-1}$-approximation for maximizing monotone submodular set functions
over a uniform matroid. This bound is now known to be tight even for a general matroid~\cite{CCPV11} whereas the greedy algorithm provides a 2-approximation for an arbitrary matroid (and a $k+1$-approximation for the intersection of
$k$ matroids) as shown in \cite{FNM78}. 
Our max-sum diversification problem with monotone submodular set functions can be viewed as an extension of that problem:
the objective function now not only contains a submodular part, but also has a super-modular part: the sum of distances. 

Since the max-sum diversification problem with modular set functions studied in~\cite{GoSh09} admits a 2-approximation algorithm, it is natural to ask what approximation ratio is obtainable for the same problem with monotone submodular set functions.
The Gollapudi and Sharma algorithm is based on the observation that the
diversity function with modular set functions can be reduced to the max-sum $p$ 
dispersion problem by changing the metric. 
Namely, the reduction defines the metric 
$d'(u,v) = w(u) + w(v) + 2 \lambda d(u,v)$. It is clear then that this
reduction and then algorithm  
in~\cite{GoSh09} does not apply 
to the submodular case where elements do not have weights but rather only
marginal weights. While this suggests that a greedy algorithm using marginal
weights might apply (as we will show), this still requires a proof and in general
one cannot expect the same approximation ratio.   
In what follows we assume (as is standard when considering submodular
functions) access to an oracle for finding an element 
$u \in U-S$ that maximizes $f(S+u)-f(S)$. When $f$ is modular, this simply 
means accessing the element $u \in U-S$ having maximum weight. 

\begin{theorem}
\label{thm:main}
There is a simple linear time greedy algorithm 
that achieves a 2-approximation for the max-sum diversification problem 
with monotone submodular set functions satisfying a cardinality constraint.
\end{theorem}

Before giving the proof 
\footnote{While greedy algorithms are conceptually simple to state
and understand operationally, it can be the case that the analysis
of an approximation ratio is not at all simple. For example, the
Birnbaum and Goldman proof that the
greedy algorithm is a 2-approximation for the cardinality constrained
metric sum dispersion problem is such a proof. Their proof answered an explicit
12 year old  conjecture by Hassin et al \cite{HassinRT97} following
the 4-approximation by Ravi et al \cite{RRT94}. In fact, one can view
the Ravi et al paper as an implicit conjecture
given their example showing that the greedy
algorithm was no better than a 2-approximation for the dispersion problem.}
 of Theorem~\ref{thm:main}, we first introduce our notation.
We extend the notion of distance function to sets. For disjoint subsets $S, T\subseteq U$, we let $d(S)=\sum_{\{u,v\}:{u,v\in S}} d(u,v)$, and $d(S, T)=\sum_{\{u,v\}:{u\in S, v\in T}} d(u,v)$. 

\medskip

Now we define various types of marginal gain.
For any given subset $S\subseteq U$ and an element $u\in U-S$: let $\phi(S)$ be the value of the objective function, $d_u(S)=\sum_{v\in S} d(u,v)$ be the marginal gain on the distance, $f_u(S)=f(S+u)-f(S)$ be the marginal gain on the weight, and $\phi_u(S)=f_u(S)+\lambda d_u(S)$ be the total marginal gain on the objective function.
Let $f'_u(S)=\frac{1}{2}f_u(S)$, and $\phi'_u(S)=f'_u(S)+\lambda d_u(S)$.
We consider the following simple greedy algorithm: 

\medskip

\noindent{\sc Greedy Algorithm}

\noindent$S=\emptyset$\\
while $|S|<p$\\
\indent \ \ \ \ \ \ find $u\in U-S$ maximizing $\phi'_u(S)$\\
\indent \ \ \ \ \ \ $S=S+u$\\
end while\\
return $S$

\medskip

Note that the above greedy algorithm is ``non-oblivious" (in the sense of~\cite{KMSV}) as it is not selecting the next element with respect to the objective function $\phi(\cdot)$. This might be of an independent interest.
We utilize the following lemma in~\cite{RRT94}.
\begin{lemma}
\label{lem:rrt}
Given a metric distance function $d(\cdot, \cdot)$, and two disjoint sets $X$ and $Y$, we have the following inequality: $(|X|-1)d(X,Y)\ge |Y|d(X).$
\end{lemma}
Now we are ready to prove Theorem~\ref{thm:main}.
\proof
Let $O$ be the optimal solution, and $G$, the greedy solution at the end of the algorithm.
Let $G_i$ be the greedy solution at the end of step $i$, $i<p$; and let $A=O\cap G_i$, $B=G_i-A$ and
$C=O-A$. By lemma~\ref{lem:rrt}, we have the following three inequalities: 
\begin{eqnarray}
(|C|-1)d(B,C)\ge |B|d(C)\\
(|C|-1)d(A,C)\ge |A|d(C)\\
(|A|-1)d(A,C)\ge |C|d(A)
\end{eqnarray}
Furthermore, we have
\begin{eqnarray}
d(A,C)+d(A)+d(C)=d(O)
\end{eqnarray}

Note that the algorithm clearly achieves the optimal solution if $p=1$.
If $|C|=1$, then $i=p-1$ and $G_i\subset O$.
Let $v$ be the element in $C$, and let $u$ be the element taken by the greedy algorithm in the next step, then
$\phi'_u(G_i) \ge \phi'_v(G_i)$. Therefore,
$ \frac{1}{2}f_u(G_i) + \lambda d_u(G_i) \ge  \frac{1}{2}f_v(G_i) + \lambda d_v(G_i),$
which implies $\phi_u(G_i) = f_u(G_i) + \lambda d_u(G_i) \ge \frac{1}{2}f_u(G_i) + \lambda d_u(G_i)$
$\ge \frac{1}{2}f_v(G_i) + \lambda d_v(G_i) \ge \frac{1}{2}\phi_v(G_i) $ and hence $\phi(G)\ge\frac{1}{2}\phi(O)$.

Now we can assume that
$p>1$ and $|C|>1$. We apply the following non-negative multipliers to equations (1), (2), (3), (4) and add them:
$(1)*\frac{1}{|C|-1}+(2)*\frac{|C|-|B|}{p(|C|-1)}+(3)*\frac{i}{p(p-1)}+(4)*\frac{i|C|}{p(p-1)}$; we then have
$d(A,C)+d(B,C)-\frac{i|C|(p-|C|)}{p(p-1)(|C|-1)}d(C)\ge \frac{i|C|}{p(p-1)}d(O).$
Since $p > |C|$,  
$d(C, G_i)\ge \frac{i|C|}{p(p-1)}d(O).$
By submodularity and monotonicity of $f'(\cdot)$, we have $\sum_{v\in C}f'_v(G_i)\ge f'(C\cup G_i)-f'(G_i)\ge  f'(O)-f'(G).$ 
Therefore, $\sum_{v\in C}\phi'_{v}(G_i) = \sum_{v\in C}[f'_v(G_i)+\lambda d(\{v\}, G_i)]$
$=\sum_{v\in C}f'_v(G_i)+\lambda d(C, G_i) \ge  [f'(O)-f'(G)]+\frac{\lambda i|C|}{p(p-1)}d(O).$

Let $u_{i+1}$ be the element taken at step $(i+1)$, then we have 
$\phi'_{u_{i+1}}(G_i)\ge \frac{1}{p}[f'(O)-f'(G)]+\frac{\lambda i}{p(p-1)}d(O).$ Summing over all $i$ from $0$ to $p-1$, we have
$\phi'(G)=\sum_{i=0}^{p-1}\phi'_{u_{i+1}}(G_i)\ge[f'(O)-f'(G)]+\frac{\lambda}{2}d(O).$
Hence, $f'(G)+\lambda d(G)\ge f'(O)-f'(G)+\frac{\lambda}{2}d(O),$ and
$\phi(G)=f(G)+\lambda d(G)\ge \frac{1}{2}[f(O)+\lambda d(O)]=\frac{1}{2}\phi(O).$
This completes the proof.
\qed

The greedy algorithm runs in time proportional to $p$ 
(for the $p$ iterations)  times the
cost of computing $\phi'_u(S)$ for a given $u$ and $S$. When $f$ is
modular, the time for updating 
$\phi'_u(S)$ can be bounded by $O(n)$. Namely, 
each iteration costs $O(n)$ time 
(to search over all elements $u$
in $U \setminus S$) and update $\phi'(S)$. 
Updating $f'(S)$ is clearly $O(1)$ while naively updating 
$d_u(S)$ would take time $O(p)$. But as observed by    
Birnbaum and Goldman \cite{BirnbaumG09},  
$d_u(V')$ can be maintained for all $V \setminus S$ within the same 
$O(n)$ needed to search $V'$ so that updating $\phi'(S)$ only costs
time $O(1)$.  
Hence the total time is $O(np)$, 
linear in $n$ when $p$ is a constant. 


\begin{corollary}
The Ravi et al. \cite{RRT94}  
greedy algorithm for dispersion has approximation
ratio no worse that 2. 
\end{corollary}

\proof 
The identically zero function $f$ is monotone submodular and 
for this $f$, our greedy algorithm is precisely the dispersion
algorithm of  Ravi et al. 
\qed

We note that for the dispersion problem, Birnbaum and Goldman \cite{BirnbaumG09}
show that their bound for the greedy
algorithm is tight. In particular, for the greedy algorithm that adds
one element at a time, the precise bound is $\frac{2p-2}{p-1}$.


%% file: matroid.tex
\section{Matroids and Local Search}
\label{sec:matroid}

Theorem~\ref{thm:main} provides a 
2-approximation for max-sum diversification when the set function is 
submodular and the set constraint is a cardinality constraint, i.e., 
a uniform matroid. It is natural to ask if the same approximation 
guarantee can be obtained for an arbitrary matroid.
In this section, we show that the max-sum diversification problem with 
monotone submodular function admits a 2-approximation 
subject to a general matroid constraint.

Matroids are well studied objects in combinatorial optimization. 
A matroid $\cal M$ is a pair $<U, {\cal F}>$, where $U$ is a set of ground elements and $\cal F$ is a collection of subsets of $U$, called {\em independent sets}, with the following properties
:
\bi
\item  {\bf Hereditary:} The empty set is independent and if $S\in {\cal F}$ and $S'\subset S$, then $S'\in {\cal F}$. 
\item  {\bf Augmentation:} If $A, B\in {\cal F}$ and $|A|>|B|$, then $\exists e\in A-B$ such that $B\cup \{e\}\in {\cal F}$.
\ei
The maximal independent sets of a matroid are called {\em bases} of $\cal M$.
Note that all bases have the same number of elements, and this number
is called the {\em rank} of $\cal M$.
The definition of a matroid captures the key notion of independence from linear algebra and extends that notion so as to apply to many combinatorial objects. 
We have already mentioned two classes of matroids relevant to our results, 
namely
partition matroids and transversal matroids. In a partition matroid, the 
universe $U$ is partitioned into sets $S_1, \ldots, S_m$ and the independent
sets $S$ satisfy  $S = \cup_{1 \leq i \leq m} S_i$ with $|S_i| \leq k_i$
for some given bounds $k_i$ on each part of the partition. 
A uniform matroid is a special case of a partition matroid with $m = 1$. 
In a transversal matroid, the universe $U$ is a union of (possibly) intersecting sets
${\cal C} = C_1, \ldots, C_m$ and a set 
$S = \{s_1, \ldots s_r\} \subseteq U$  is independent if there is an
injective function $\phi$
from $S$ into ${\cal C}$ with say $\phi(s_i) = C_i$ and  $\phi(s_i) \in C_i$.That is, $S$ forms a set of representatives for each set $C_i$ or equivalently 
there is a matching between $S$ and ${\cal C}$. 
(Note that a given $s_i$ could occur in other sets $C_j$.)   

\medskip
 
{\sc Problem 2.} {\tt Max-Sum Diversification for Matroids} \\ 
\\
Let $U$ be the underlying ground set, and $\cal F$ be the set of independent subsets of $U$ such that ${\cal M}=<U, {\cal F}>$ is a matroid.
Let $d(\cdot,\cdot)$ be a (non-negative) metric distance function measuring the distance on every pair of elements. For any subset of $U$, let $f(\cdot)$ be a non-negative monotone submodular set function measuring the weight of the subset. The goal of the problem is to find a subset $S \in {\cal F}$ that:
$$\begin{array}{ll}
\rm{maximizes} & f(S)+\lambda\sum_{\{u,v\}:u,v\in S} d(u,v)
\end{array}$$
where $\lambda$ is a parameter specifying a desired trade-off between the two objectives.
As before, we let $\phi(S)$ be the value of the objective function. 
Note that since the function $\phi(\cdot)$ is monotone, 
$S$ is essentially a basis of the matroid ${\cal M}$.
The greedy algorithm in Section~\ref{sec:submo} still applies, but it fails to achieve any constant approximation ratio even for a linear quality function 
$fcdot)$ including the identically zero function; that is, for 
max-sum dispersion. (See the Appendix.) 
This is in contrast to the seminal result of Nemhauser, Wolsey and Fisher 
\cite{NWF78} showng that the greedy algorithm is optimal (respectivley, 
a 2-approximation) 
for linear functions (respectively, monotone submodular functions) subject to a matroid constraint.


Note that the problem is trivial if the rank of the matroid is less than two.
Therefore, without loss of generality, we assume the rank is greater or equal to two.
Let $$\{x,y\}=\argmax_{\{x,y\}\in {\cal F}}[f(\{x,y\})+\lambda d(x,y)].$$ 
We now consider the following oblivious local search algorithm: 

\medskip

\noindent{\sc Local Search Algorithm}

\noindent let $S$ be a basis of $\cal M$ containing both $x$ and $y$\\
while there is an $u\in U-S$ and $v\in S$ such that $S+u-v\in {\cal F}$ and $\phi(S+u-v) >\phi(S)$ \\
\indent \ \ \ \ \ \ $S=S+u-v$\\
end while\\
return $S$

\medskip

\begin{theorem} 
\label{thm:ls}
The local search algorithm achieves an approximation ratio of 2 for max-sum diversification with a matroid constraint.
\end{theorem}

Note that if the rank of the matroid is two, then the algorithm is clearly optimal. 
From now on, we assume the rank of the matroid is greater than two. 
Before we prove the theorem, we first give several lemmas.
All the lemmas assume the problem and the underlying matroid without explicitly mentioning it.
Let $O$ be the optimal solution, and $S$, the solution at the end of the local search algorithm.
Let $A=O\cap S$, $B=S-A$ and $C=O-A$.
\begin{lemma} 
\label{lem:bj}
For any two sets $X, Y\in {\cal F}$ with $|X|=|Y|$, there is a bijective mapping $g: X\rightarrow Y$ such that $X-x+g(x)\in {\cal F}$ for any $x\in X$.
\end{lemma}

This is a known property of a matriod and its proof can be found in~\cite{Brualdi-1969}.
Since both $S$ and $O$ are bases of the matroid, they have the same cardinality.
Therefore, $B$ and $C$ have the same cardinality.
By Lemma~\ref{lem:bj}, there is a bijective mapping $g: B\rightarrow C$ such that $S-b+g(b)\in {\cal F}$ for any $b\in B$.
Let $B=\{b_1, b_2, \dots, b_t\}$, and let $c_i=g(b_i)$ for all $i$.
Without loss of generality, we assume $t\ge 2$, for otherwise, the algorithm is optimal by the local optimality condition.
\begin{lemma} 
\label{lem:sm1}
$f(S)+\sum_{i=1}^{t} f(S-b_i+c_i)\ge f(S-\sum_{i=1}^{t} b_i)+\sum_{i=1}^{t} f(S+c_i)$.
\end{lemma}
\proof
Since $f$ is submodular, 
$$f(S)-f(S-b_1)\ge f(S+c_1)-f(S+c_1-b_1)$$
$$f(S-b_1)-f(S-b_1-b_2)\ge f(S+c_2)-f(S+c_2-b_2)$$
$$\vdots$$
$$f(S-\sum_{i=1}^{t-1}b_i)-f(S-\sum_{i=1}^{t}b_i)\ge f(S+c_t)-f(S+c_t-b_t).$$
Summing up these inequalities, we have
$$f(S)- f(S-\sum_{i=1}^{t} b_i)\ge \sum_{i=1}^{t} f(S+c_i) - \sum_{i=1}^{t} f(S-b_i+c_i),$$
and the lemma follows.
\qed

\begin{lemma} 
\label{lem:sm2}
$\sum_{i=1}^t f(S+c_i)\ge (t-1)f(S)+f(S+\sum_{i=1}^t c_i)$.
\end{lemma}
\proof
Since $f$ is submodular, 
$$f(S+c_t)-f(S)= f(S+c_t)-f(S)$$
$$f(S+c_{t-1})-f(S)\ge f(S+c_t+c_{t-1})-f(S+c_t)$$
$$f(S+c_{t-2})-f(S)\ge f(S+c_t+c_{t-1}+c_{t-2})-f(S+c_t+c_{t-1})$$
$$\vdots$$
$$f(S+c_1)-f(S)\ge f(S+\sum_{i=1}^t c_i)-f(S+\sum_{i=2}^t c_i)$$
Summing up these inequalities, we have
$$\sum_{i=1}^t f(S+c_i) - tf(S)\ge f(S+\sum_{i=1}^t c_i)-f(S),$$
and the lemma follows.
\qed

\begin{lemma} 
\label{lem:sm3}
$\sum_{i=1}^{t} f(S-b_i+c_i)\ge (t-2)f(S)+f(O)$.
\end{lemma}
\proof
Combining Lemma~\ref{lem:sm1} and Lemma~\ref{lem:sm2}, we have
\begin{eqnarray*}
& &f(S)+ \sum_{i=1}^{t} f(S-b_i+c_i)\\
&\ge&f(S-\sum_{i=1}^{t} b_i)+\sum_{i=1}^{t} f(S+c_i)\\
&\ge&(t-1)f(S)+f(S+\sum_{i=1}^{t} c_i)\\
&=&(t-1)f(S)+f(S+C)\\
&\ge&(t-1)f(S)+f(O).
\end{eqnarray*}
Therefore, the lemma follows.
\qed

\begin{lemma} 
\label{lem:eg1}
If $t>2$, $d(B,C) - \sum_{i=1}^{t} d(b_i, c_i)\ge d(C)$.
\end{lemma}
\proof
For any $b_i, c_j, c_k$, we have
$$d(b_i, c_j)+d(b_i,c_k)\ge d(c_j, c_k).$$
Summing up these inequalities over all $i,j,k$ with $i\neq j$, $i\neq k$, $j\neq k$, we have
each $d(b_i, c_j)$ with $i\ne j$ is counted $(t-2)$ times; and each $d(c_i, c_j)$ with $i\ne j$ is counted $(t-2)$ times.
Therefore $$(t-2)[d(B,C) - \sum_{i=1}^{t} d(b_i, c_i)]\ge (t-2)d(C),$$ and the lemma follows.
\qed

\begin{lemma} 
\label{lem:eg2}
$\sum_{i=1}^t d(S-b_i+c_i)\ge (t-2)d(S)+d(O)$.
\end{lemma}
\proof
\begin{eqnarray*}
& &\sum_{i=1}^t d(S-b_i+c_i)\\
&=&\sum_{i=1}^t [d(S)+d(c_i, S-b_i)-d(b_i, S-b_i)]\\
&=& td(S)+\sum_{i=1}^{t}d(c_i, S-b_i)-\sum_{i=1}^{t}d(b_i, S-b_i)\\
&=& td(S)+\sum_{i=1}^{t}d(c_i, S)-\sum_{i=1}^{t}d(c_i,b_i)-\sum_{i=1}^{t}d(b_i, S-b_i)\\
&=& td(S)+d(C, S)-\sum_{i=1}^{t}d(c_i,b_i)-d(A, B)-2d(B).
\end{eqnarray*}
There are two cases.
If $t>2$ then by Lemma~\ref{lem:eg2}, we have
\begin{eqnarray*}
& &d(C,S)-\sum_{i=1}^{t}d(c_i,b_i)\\
&=&d(A,C)+d(B,C)-\sum_{i=1}^{t}d(c_i,b_i)\\
&\ge&d(A,C)+d(C).
\end{eqnarray*}
Furthermore, since $d(S)=d(A)+d(B)+d(A,B)$,
we have $2d(S)-d(A,B)-2d(B)\ge d(A)$.
Therefore
\begin{eqnarray*}
& &\sum_{i=1}^t d(S-b_i+c_i)\\
&=& td(S)+d(C, S)-\sum_{i=1}^{t}d(c_i,b_i)-d(A, B)-2d(B)\\
&\ge&  (t-2)d(S)+d(A,C)+d(C)+d(A)\\
&\ge&  (t-2)d(S)+d(O).
\end{eqnarray*}
If $t=2$, then since the rank of the matroid is greater than two, $A\neq\emptyset$. 
Let $z$ be an element in $A$, then we have
\begin{eqnarray*}
& & 2d(S)+d(C, S)-\sum_{i=1}^{t}d(c_i,b_i)-d(A, B)-2d(B)\\
&=&d(A,C)+d(B,C)-\sum_{i=1}^{t}d(c_i,b_i)+2d(A)+d(A,B)\\
&\ge&d(A,C)+d(c_1,b_2)+d(c_2,b_1)+d(A)+d(z,b_1)+d(z,b_2)\\
&\ge&d(A,C)+d(A)+d(c_1, c_2)\\
&\ge&d(A,C)+d(A)+d(C)\\
&=&d(O).
\end{eqnarray*}
Therefore
\begin{eqnarray*}
& &\sum_{i=1}^t d(S-b_i+c_i)\\
&=& td(S)+d(C, S)-\sum_{i=1}^{t}d(c_i,b_i)-d(A, B)-2d(B)\\
&\ge&  (t-2)d(S)+d(O).
\end{eqnarray*}
This completes the proof.
\qed

Now with the proofs of Lemma~\ref{lem:sm3} and Lemma~\ref{lem:eg2},
we are ready to complete the proof of Theorem~\ref{thm:ls}.
\proof
Since $S$ is a locally optimal solution, we have 
$\phi(S)\ge\phi(S-b_i+c_i)$ for all $i$.
Therefore, for all $i$ we have
$$f(S)+\lambda d(S)\ge f(S-b_i+c_i)+\lambda d(S-b_i+c_i).$$
Summing up over all $i$, we have
$$tf(S)+\lambda td(S)\ge \sum_{i=1}^{t}f(S-b_i+c_i)+\lambda \sum_{i=1}^{t}d(S-b_i+c_i).$$
By Lemma~\ref{lem:sm3}, we have
$$tf(S)+\lambda td(S)\ge (t-2)f(S)+f(O)+\lambda \sum_{i=1}^{t}d(S-b_i+c_i).$$
By Lemma~\ref{lem:eg2}, we have
$$tf(S)+\lambda td(S)\ge (t-2)f(S)+f(O)+\lambda [(t-2)d(S)+d(O)].$$
Therefore,
$$2f(S)+2\lambda d(S))\ge f(O)+\lambda d(O).$$
$$\phi(S)\ge\frac{1}{2}\phi(O),$$
this completes the proof.
\qed

Theorem~\ref{thm:ls} shows that even in the more general case of a matroid constraint,
we can still achieve the approximation ratio of 2. 
As is standard in such local search algorithms, 
with a small sacrifice on the approximation ratio,
the algorithm can be modified to run in polynomial time by requiring at least an  
$\epsilon$-improvement at each iteration rather than just any
improvement.

%% file: dynup.tex
\section{Dynamic Update}
\label{sec:dynup}
In this section, we discuss dynamic updates for the max-sum diversification problem with modular set functions. 
The setting is that we have initially computed a good solution with some approximation guarantee. The weights are changing over time, and upon seeing a change of weight, we want to maintain the quality (the same approximation ratio) of the solution by modifying the current solution without completely recomputing it. 
We use the number of updates to quantify the amount of modification needed to
maintain the desired approximation.
An {\em update} is a single swap of an element in $S$ with an element outside $S$, where $S$ is the current solution. We ask the following question:
\begin{quote}
Can we maintain a good approximation ratio with a limited number of updates?
\end{quote}
Since the best known approximation algorithm achieves approximation ratio of 2, it is natural to ask whether it is possible to maintain that ratio through local updates. 
And if it is possible, how many such updates it requires.
To simplify the analysis, we restrict to the following oblivious update rule.
Let $S$ be the current solution, and let $u$ be an element in $S$ and $v$ be an element outside $S$. The marginal gain
$v$ has over $u$ with respect to $S$ is defined to be
$$\phi_{v\rightarrow u}(S)=\phi(S\setminus\{u\}\cup\{v\})-\phi(S).$$

\noindent{\sc Oblivious (single element swap) Update Rule}\\
Find a pair of elements $(u,v)$ with $u\in S$ and $v\not\in S$ maximizing $\phi_{v\rightarrow u}(S)$.
If $\phi_{v\rightarrow u}(S)\le 0$, do nothing; otherwise swap $u$ with $v$. 

Since the oblivious local search in Theorem~\ref{thm:ls} uses the same
single element swap update rule, it is not hard to see that we can maintain 
the approximation ratio of 2.
However, it is not clear how many updates 
are needed to maintain that ratio. We conjecture that the number of updates
can be made relatively small (i.e., constant) by a non-oblivious update rule and carefully maintaining some desired configuration of the solution set. We leave this as an open question.

However, we are able to show that if we relax the requirement slightly, i.e.,
aiming for an approximation ratio of 3 instead of 2, and restrict slightly the magnitude of the weight-perturbation,
we are able to maintain the desired ratio with a single update.
Note that the weight restriction is only used for the case of a weight decrease 
(Theorem \ref{thm:wd}). 

We divide weight-perturbations into four types: a weight increase (decrease) which occurs on an element,
and a distance increase (decrease) which occurs between two elements. 
We denote these four types: {\sc (i), (ii),(iii), (iv)}; and we have a corresponding theorem for each case.


Before getting to the theorems, we first prove the following two lemmas.
After a weight-perturbation, let $S$ be the 
current solution set, and $O$ be the
optimal solution.
Let $S^*$ be the solution set after a single update using the oblivious update rule,
and let $\Delta=\phi(S^*)-\phi(S)$.
We again let $Z=O\cap S$, $X=O\setminus S$ and $Y=S\setminus O$. 
\begin{lemma}
\label{lem:up}
There exists $z\in Y$ such that $$\phi_z(S\setminus\{z\})\le\frac{1}{|Y|}[f(Y)+2\lambda d(Y)+\lambda d(Z,Y)].$$
\end{lemma}
\proof
If we sum up all marginal gain $\phi_y(S\setminus\{y\})$ for all $y\in Y$, we have
$$\sum_{y\in Y}\phi_y(S\setminus\{y\}) = f(Y)+2\lambda d(Y)+\lambda d(Z,Y).$$
By 
an averaging argument, there must exist $z\in Y$ such that 
$$\phi_z(S\setminus\{z\})\le\frac{1}{|Y|}[f(Y)+2\lambda d(Y)+\lambda d(Z,Y)].$$
\qed

Lemma~\ref{lem:up} ensures the existence of an element in $S$
such that after removing it from $S$, the objective function value does not decrease much.
The following lemma ensures that there always exists
an element outside $S$ which can increase the objective function value substantially if we bring it in.

\begin{lemma}
\label{lem:low}
If $\phi(S^*)< \frac{1}{3}\phi(O)$, then for all $y\in Y$, there exists $x\in X$ such that 
$$\phi_x(S\setminus\{y\})>\frac{1}{|X|}[2\phi(Z)+3\phi(Y)+3\lambda d(Z,Y)+3\Delta].$$
\end{lemma}
\proof
For any $y\in Y$, and by Lemma~\ref{lem:rrt}, we have 
\begin{eqnarray*}
&&f(X)+\lambda d(S\setminus\{y\}, X)\\
&=&f(X)+\lambda d(Z,X)+\lambda d(Y\setminus\{y\}, X)\\
&\ge& f(X)+\lambda d(Z,X)+\lambda d(X).
\end{eqnarray*}
Note that since $\phi(S^*)=\phi(S)+\Delta< \frac{1}{3}\phi(O)$, we have
\begin{eqnarray*}
\phi(O)&=&\phi(Z)+f(X)+\lambda d(X)+\lambda d(Z,X)\\
&>&3\phi(Z)+3\phi(Y)+3\lambda d(Z,Y)+3\Delta.
\end{eqnarray*}
Therefore,
\begin{eqnarray*}
&&f(X)+\lambda d(S\setminus\{y\}, X)\\
&\ge& f(X)+\lambda d(Z,X)+\lambda d(X)\\
&>&2\phi(Z)+3\phi(Y)+3\lambda d(Z,Y)+3\Delta.
\end{eqnarray*}
This implies there must exist $x\in X$ such that 
$$\phi_x(S\setminus\{y\})>\frac{1}{|X|}[2\phi(Z)+3\phi(Y)+3\lambda d(Z,Y)+3\Delta].$$
\qed

Combining Lemma~\ref{lem:up} and \ref{lem:low}, we can give a lower bound for $\Delta$.
We have the following corollary.
\begin{corollary}
\label{cor:delta}
If $\phi(S^*)< \frac{1}{3}\phi(O)$, then we have $|Y|> 3$ and furthermore
$$\Delta>\frac{1}{|Y|-3}[2\phi(Z)+2f(Y)+\lambda d(Y)+2\lambda d(Z,Y)].$$
\end{corollary}
\proof
By Lemma~\ref{lem:up}, there exists $y\in Y$ such that $$\phi_y(S\setminus\{y\})\le\frac{1}{|Y|}[f(Y)+2\lambda d(Y)+\lambda d(Z,Y)].$$
Since $\phi(S^*)< \frac{1}{3}\phi(O)$,
by Lemma~\ref{lem:low}, for this particular $y$, there exists $x\in X$ such that 
$$\phi_x(S\setminus\{y\})>\frac{1}{|X|}[2\phi(Z)+3\phi(Y)+3\lambda d(Z,Y)+3\Delta].$$
Since $|X|=|Y|$, we have 
$$\Delta>\frac{1}{|Y|}[2\phi(Z)+2f(Y)+\lambda d(Y)+2\lambda d(Z,Y)+3\Delta].$$
If $|Y|\le 3$, then it is a contradiction. Therefore $|Y|> 3$. 
Rearranging the inequality, we have
$$\Delta>\frac{1}{|Y|-3}[2\phi(Z)+2f(Y)+\lambda d(Y)+2\lambda d(Z,Y)].$$
\qed

\begin{corollary}
\label{cor:small-p}
If $p \leq 3$, then for any weight or distance perturbation, we
can maintain an approximation ratio of 3 with a single update.
\end{corollary}

\proof
This is an immediate consequence of Corollary~\ref{cor:delta} since
$p \geq |Y|$.
\qed

Given Corollary~\ref{cor:small-p}, we will assume $p > 3$ for
all the remaining results in this section. We first discuss weight-perturbations on elements.
\begin{theorem}
\label{thm:wi}
{\sc [type (i)]}
For any weight increase, we can maintain an approximation ratio of 3 with a single update.
\end{theorem}
\proof
Suppose we increase the weight of $s$ by $\delta$. 
Since the optimal solution can increase by at most $\delta$, if $\Delta \geq \frac{1}{3} \delta$,
then we have maintained a ratio of 3. Hence we assume
$\Delta < \frac{1}{3} \delta$.
If $s\in S$ or $s\not\in O$, then it is clear the ratio of $3$ is maintained. The only interesting case is when $s\in O\setminus S$. Suppose, for the sake of contradiction, that $\phi(S^*)<\frac{1}{3}\phi(O)$, then by Corollary~\ref{cor:delta}, we have $|Y|> 3$ and
$$\Delta>\frac{1}{|Y|-3}[2\phi(Z)+2f(Y)+\lambda d(Y)+2\lambda d(Z,Y)].$$
Since $\Delta<\frac{1}{3}\delta$, we have
$$\delta>\frac{1}{|Y|-3}[6\phi(Z)+6f(Y)+3\lambda d(Y)+6\lambda d(Z,Y)].$$
On the other hand, by Lemma~\ref{lem:up}, there exists $y\in Y$ such that $$\phi_y(S\setminus\{y\})\le\frac{1}{|Y|}[f(Y)+2\lambda d(Y)+\lambda d(Z,Y)].$$
Now considering a swap of s with y, the loss by removing $y$ from $S$ is
$\phi_y(S\setminus\{y\})$, while the increase that $s$ brings to the set $S\setminus\{y\}$ is at
least $\delta$ (as $s$ is increased by $\delta$, and the original weight of
$s$ is non-negative). Therefore the marginal gain of the swap of $s$ with $y$ is
$\phi_{s \rightarrow y} \geq \delta - \phi_y(S\setminus\{y\})$ and hence
$$\phi_{s\rightarrow y}(S)\ge\delta-\frac{1}{|Y|}[f(Y)+2\lambda d(Y)+\lambda d(Z,Y)].$$
However,  $\phi_{s\rightarrow y}(S)\le\Delta<\frac{1}{3}\delta$. Therefore, we have
$$\frac{1}{3}\delta>\delta-\frac{1}{|Y|}[f(Y)+2\lambda d(Y)+\lambda d(Z,Y)].$$
This implies
$$\delta<\frac{1}{|Y|}[\frac{3}{2}f(Y)+3\lambda d(Y)+\frac{3\lambda}{2}d(Z,Y)],$$
which is a contradiction. 
\qed

\begin{theorem}
\label{thm:wd}
{\sc [type (ii)]}
For a weight decrease of magnitude $\delta$, 
we can maintain an approximation ratio of 3 with
$$\lceil\log_{\frac{p-2}{p-3}}\frac{w}{w-\delta}\rceil$$ 
updates, where $w$ is the weight of the solution before the weight decrease.
In particular, if $\delta\le \frac{w}{p-2}$, we only need a single update.
\end{theorem}
\proof
Suppose we decrease the weight of $s$ by $\delta$. Without loss of generality, we can assume $s\in S$.
Suppose, for the sake of contradiction, that $\phi(S^*)<\frac{1}{3}\phi(O)$, then 
by Corollary~\ref{cor:delta}, we have $|Y|> 3$ and
\begin{eqnarray*}
\Delta&>&\frac{1}{|Y|-3}[2\phi(Z)+2f(Y)+\lambda d(Y)+2\lambda d(Z,Y)]\\
&\ge&\frac{1}{p-3}\phi(S).
\end{eqnarray*}
Therefore $$\phi(S^*)>\frac{p-2}{p-3}\phi(S).$$
This implies that we can maintain the approximation ratio with 
$$\lceil\log_{\frac{p-2}{p-3}}\frac{w}{w-\delta}\rceil$$ number of updates.
In particular, if $\delta\le \frac{w}{p-2}$, we only need a single update.
\qed

We now discuss the weight-perturbations between two elements. We assume that such perturbations preserve the metric condition.
Furthermore, we assume $p>3$ for otherwise, by Corollary~\ref{cor:delta}, the ratio of 3 is maintained.
\begin{theorem}
\label{thm:di}
{\sc [type (iii)]}
For any distance increase, we can maintain an approximation ratio of 3 with a single update.
\end{theorem}
\proof
Suppose we increase the distance of $(x,y)$ by $\delta$,
and for the sake of contradiction, we assume that $\phi(S^*)<\frac{1}{3}\phi(O)$, then 
by Corollary~\ref{cor:delta}, we have $|Y|> 3$ and
$$\Delta>\frac{1}{|Y|-3}[2\phi(Z)+2f(Y)+\lambda d(Y)+2\lambda d(Z,Y)].$$
Since $\Delta<\frac{1}{3}\delta$, we have
\begin{eqnarray*}
\delta&>&\frac{3}{|Y|-3}[2\phi(Z)+2f(Y)+\lambda d(Y)+2\lambda d(Z,Y)]\\
&\ge&\frac{3}{p-3}\phi(S).
\end{eqnarray*}
If both $x$ and $y$ are in $S$, then it is not hard to see that the ratio of $3$ is maintained. Otherwise, 
there are two cases:
\be
\item 
Exactly one of $x$ and $y$ is in $S$, without loss of generality, we assume $y\in S$. 
Considering that we swap $x$ with any vertex $z\in S$ other than $y$. Since after the swap, both $x$ and $y$ are now in $S$, by the triangle inequality of the metric condition, we have 
$$\Delta\ge(p-1)\delta-\phi(S)>(\frac{2}{3}p-2)\delta.$$
Since $p>3$, we have
$$\Delta>(\frac{2}{3}p-2)\delta\ge\frac{2}{3}\delta>2\Delta,$$
which is a contradiction.
\item
Both $x$ and $y$ are outside in $S$.
By Lemma~\ref{lem:up},
there exists $z\in Y$ such that $$\phi_z(S\setminus\{z\})\le\frac{1}{|Y|}[f(Y)+2\lambda d(Y)+\lambda d(Z,Y)].$$
Consider the set $T=\{x,y\}$ with $S\setminus\{z\}$, by the triangle inequality of the metric condition, we have $d(T,S\setminus\{z\})\ge (p-1)\delta$. Therefore, at least one of $x$ and $y$, without loss of generality, assuming $x$, has the following property:
$$d(x,S\setminus\{z\})\ge\frac{(p-1)\delta}{2}.$$
Considering that we swap $x$ with $z$, we have: 
$$\Delta\ge\frac{(p-1)}{2}\delta-\frac{1}{|Y|}[f(Y)+2\lambda d(Y)+\lambda d(Z,Y)].$$
Since $\Delta<\frac{1}{3}\delta$, we have
$$\frac{1}{3}\delta>\frac{(p-1)}{2}\delta-\frac{1}{|Y|}[f(Y)+2\lambda d(Y)+\lambda d(Z,Y)].$$
This implies that
$$\delta<\frac{6}{3p-5}\cdot\frac{1}{|Y|}[f(Y)+2\lambda d(Y)+\lambda d(Z,Y)].$$
Since $p>3$, we have
$$\delta<\frac{1}{|Y|}[\frac{6}{7}f(Y)+\frac{12\lambda}{7}d(Y)+\frac{6\lambda}{7}d(Z,Y)],$$
which is a contradiction.
\ee
Therefore, $\phi(S^*)\ge\frac{1}{3}\phi(O)$; this completes the proof.
\qed

\begin{theorem}
\label{thm:de}
{\sc [type (iv)]}
For any distance decrease, we can maintain an approximation ratio of 3 with a single update.
\end{theorem}
\proof
Suppose we decrease the distance of $(x,y)$ by $\delta$. Without loss of generality,
we assume both $x$ and $y$ are in $S$, for otherwise, it is not hard to see the ratio
of 3 is maintained.
Suppose, for the sake of contradiction, that $\phi(S^*)<\frac{1}{3}\phi(O)$, then 
by Corollary~\ref{cor:delta}, we have $|Y|> 3$ and
\begin{eqnarray*}
\Delta&>&\frac{1}{|Y|-3}[2\phi(Z)+2f(Y)+\lambda d(Y)+2\lambda d(Z,Y)]\\
&\ge&\frac{1}{p-3}\phi(S).
\end{eqnarray*}
If $\Delta\ge\delta$, then the ratio of 3 is maintained.
Otherwise, $$\delta>\Delta\ge\frac{1}{p-3}\phi(S).$$
By the triangle inequality of the metric condition, we have 
$$\phi(S)\ge(p-2)\delta>\frac{p-2}{p-3}\phi(S)>\phi(S),$$
which is a contradiction. 
\qed

Combining Theorem~\ref{thm:wi}, \ref{thm:wd}, \ref{thm:di}, \ref{thm:de},
we have the following corollary.
\begin{corollary}
\label{cor:main}
If the initial solution achieves approximation ratio of 3,
then for any weight-perturbation of {\sc type (i), (iii), (iv)};
and any weight-perturbation of {\sc type (ii)} that is no more than
$\frac{1}{p-2}$ of the current solution for $p>3$ and arbitrary for $p\le 3$,
we can maintain the ratio of 3 with a single update. 
\end{corollary}

%% file: experiments_tkde.tex
\begin{table*}[t]
\caption{Comparison of Greedy A and Greedy B ($N=50$)}
\centering
\begin{tabular}{|c|c|c|c|c|c|c|c|} 
\hline
$p$ & $OPT$ & $Greedy A$ & $Greedy B$ & $AF_{Greedy A}$ & $AF_{Greedy B}$ & $AF_{\frac{Greedy B}{Greedy A}}$ \\ 
\hline
3	& 4.870	& 4.311	& 4.785	& 1.130	& 1.018	& 1.110 \\
4	& 7.822	& 7.431	& 7.616	& 1.053	& 1.027	& 1.025 \\
5	& 11.202	& 10.391	& 10.933	& 1.078	& 1.025	& 1.052 \\
6	& 15.221	& 14.467	& 14.891	& 1.052	& 1.022	& 1.029 \\
7	& 11.079	& 10.178	& 10.854	& 1.089	& 1.021	& 1.066 \\
\hline
\end{tabular}
\label{table:1}
\end{table*}

\begin{table*}[t]
\caption{Comparison of Greedy A, Greedy B and LS ($N=500$)}
\centering
\begin{tabular}{|c|c|c|c|c|c|c|c|c|c|c|} 
\hline
$p$ & $Greedy A$ & $Greedy B$ & $LS$
& $AF_{\frac{Greedy B}{Greedy A}}$  & $AF_{\frac{LS}{Greedy B}}$ & $Time_{Greedy A}$ & $Time_{Greedy B}$ & $Time_{\frac{Greedy A}{Greedy B}}$\\ 
\hline
5	& 10.811	& 11.370	&11.766	&1.052	& 1.035	&4713 ms &	24 ms	& 196.375 \\
10	& 37.781	& 38.243	& 39.750	& 1.012	& 1.039	& 5934 ms & 	58 ms	& 102.310 \\
15	& 77.371	& 81.095	& 83.090	& 1.048	& 1.025	& 7255 ms & 	93 ms	& 78.011 \\
20	& 134.399	& 137.769	 & 138.900	& 1.025	& 1.008	& 9317 ms	 & 194 ms	& 48.026 \\
25	& 204.996	& 210.220	& 212.200	& 1.025	& 1.009	& 10300 ms & 	342 ms	& 30.117 \\
30	& 291.165	& 296.798	& 297.660	& 1.019	& 1.003	& 12506 ms & 	571 ms	& 21.902 \\
35	& 392.604	& 401.210	& 400.120	& 1.022	& 1.000	& 13035 ms & 	762 ms      & 17.106 \\
40	& 507.944	& 517.275	& 519.850	& 1.018	& 1.005	& 14853 ms & 	989 ms	& 15.018 \\
45	& 635.792	& 650.850	& 649.400	& 1.024	& 1.000	& 15225 ms & 	1200 ms	& 12.688 \\
50	& 782.639	& 799.470	& 802.400	& 1.022	& 1.004	& 16468 ms & 	1379 ms    & 11.942 \\
55	& 944.030	& 960.750	& 959.550	& 1.018	& 1.000	& 18150 ms & 	1615 ms	& 11.238 \\
60	& 1121.710 &	1137.800	& 1136.800	& 1.014	& 1.000	 & 18901 ms & 	1680 ms	 & 11.251  \\
65	& 1308.511 &	1332.390	& 1333.630	& 1.018	& 1.001	 & 19648 ms & 	2022 ms	 & 9.717 \\
70	& 1515.522 &	1538.470	& 1540.500	& 1.015	& 1.001	 & 19432 ms & 	2393 ms	 & 8.120 \\
75	& 1734.725 &	1761.480	& 1760.540	& 1.015	& 1.000	 & 21915 ms & 	2941 ms	 & 7.452 \\
\hline
\end{tabular}
\label{table:2}
\end{table*}

\begin{table*}[t]
\caption{Comparison of Improved Greedy A and Improved Greedy B ($N=50$)}
\centering
\begin{tabular}{|c|c|c|c|c|c|c|c|} 
\hline
$p$ & $OPT$ & $Greedy A$ & $Greedy B$ & $AF_{Greedy A}$ & $AF_{Greedy B}$ & $AF_{\frac{Greedy B}{Greedy A}}$ \\ 
\hline
3 &	4.858	& 4.754	& 4.858	& 1.022	& 1.000	& 1.022 \\
4 &	7.777	& 7.495	& 7.319	& 1.038	& 1.063	& 0.977 \\
5 &	11.013	& 10.698	& 10.885	& 1.029	& 1.012	& 1.017 \\
6 &	15.487	& 14.734	& 15.089	& 1.051	& 1.026	& 1.024 \\
7 &	19.845	& 19.264	& 19.498	& 1.030	& 1.018	& 1.012 \\
\hline
\end{tabular}
\label{table:3}
\end{table*}

\section{Experiments}
\label{sec:expmt}
While the results in this paper are mainly theoretical in nature, 
we present some  
experimental results in
this section to provide additional insight 
about the relative performance and efficiency of our algorithms. 
In section \ref{synthetic-greedy}, we will first consider the relative 
performance of two greedy algorithms and local search with respect to 
a synthetic 
data set, followed in section \ref{letor-greedy} 
by similar experiments for a 
well-known dataset (LETOR)  that has been actively used for different information 
and machine learning problems and especially for "learn to rank" research \cite{Qin-etal10}. In section \ref{dynamic-experiments}, we again consider the 
synthetic data set as in section \ref{synthetic-greedy} and make some 
obervations on the performance of local search for dynamically changing data.   

For the synthetic data as well as the LETOR data set, we consider the max-sum diversification problem with modular set functions
and a cardinality constraint $p$ so as to be able to compare the greedy
and local search algorithms as well as comparing our greedy algorithm with
the algorithm 
 of Gollapudi and Sharma \cite{GoSh09} whose work motivated this paper.
We will refer to their diversification algorithm as Greedy A. We recall
that their algorithm
consists of a reduction to the max-sum p-dispersion problem and then uses the
Hassin, Rubenstein and Tamir \cite{HassinRT97}
algorithm that greedily chooses {\it edges}
yielding an approximation ratio of $2$. We will experimentally compare the performance and
time complexity of their algorithm against our greedy by vertices algorithm
which also has approximation ratio 2. We will refer to our greedy algorithm
as Greedy B. We also consider how much a limited amount of 
local search improves the results obtained by our Greedy B algorithm. That is, we follow Greedy B by a 1-swap local
search algorithm that looks for the best improvement in each iteration.
We refer to this local search algorithm as LS with the understanding that it is
being initialized by Greedy B and terminated when either a local maximum is 
reached or when the algorithm runs for 
ten times the time of the Greedy B initialization. 

\begin{table*}[t]
\caption{Comparison of Greedy A and Greedy B ($N=50$)}
\centering
\begin{tabular}{|c|c|c|c|c|c|c|c|} 
\hline
$p$ & $OPT$ & $Greedy A$ & $Greedy B$ & $AF_{Greedy A}$ & $AF_{Greedy B}$ & $AF_{\frac{Greedy B}{Greedy A}}$ \\ 
\hline
3	& 7.088	& 6.140	& 7.088	& 1.154	& 1.000	& 1.154 \\
4	& 10.02	& 10.020	& 10.000	& 1.000	& 1.002	& 0.998 \\
5	& 12.571	& 12.470	& 12.570	& 1.008	& 1.000	& 1.008 \\
6	& 15.315	& 15.060	& 15.060	& 1.017	& 1.017	& 1.000 \\
7	& 18.54	& 17.290	& 17.949	& 1.072	& 1.033	& 1.038 \\
\hline
\end{tabular}
\label{table:4}
\end{table*}

\begin{table*}[t]
\caption{Comparison of Greedy A, Greedy B and LS}
\centering
\begin{tabular}{|c|c|c|c|c|c|c|c|c|} 
\hline
$p$ & $Greedy A$ & $Greedy B$ & $LS$
& $AF_{\frac{Greedy B}{Greedy A}}$  & $AF_{\frac{LS}{Greedy B}}$ & $Time_{Greedy A}$ & $Time_{Greedy B}$ & $Time_{\frac{Greedy A}{Greedy B}}$\\ 
\hline
5	& 13.996	& 13.999	& 	13.999	& 	1.000	& 	1.000	& 	2365 ms	& 	426 ms	& 	5.552 \\
10	& 37.570	& 37.970	& 	37.970	& 	1.011		& 1.000		& 2370 ms	& 	504 ms	& 	4.702 \\
15	& 69.590	& 71.600	& 	71.600	& 	1.029	& 	1.000	& 	2694 ms	& 	421 ms	& 	6.399 \\
20	& 110.900	& 113.640	& 	113.640		& 1.025	& 	1.000	& 	3280 ms	& 	470 ms	& 	6.979 \\
25	& 154.590	& 162.400	& 	162.480		& 1.051	& 	1.000	& 	3223	ms & 	587 ms	& 	5.491 \\
30	& 192.260	& 220.450	& 	220.730	& 	1.147	& 	1.001	& 	4364	ms	& 785 ms	& 	5.559 \\
35	& 253.790	& 288.490	& 	288.970	& 	1.137	& 	1.002	& 	4762	ms	& 758 ms	& 	6.282 \\
40	& 317.290	& 366.520	& 	367.215	& 	1.155	& 	1.002	& 	4599	ms	& 864 ms	& 	5.323 \\
45	& 397.230	& 454.500	& 	455.100	& 	1.144	& 	1.001	& 	6088	ms	& 1028 ms	& 	5.922 \\
50	& 486.440	& 552.500	& 	553.150	& 	1.136	& 	1.001	& 	5323	ms	& 1155 ms	& 	4.609 \\
55	& 584.830	& 660.430	& 	661.370	& 	1.129	& 	1.001	& 	7360	ms	& 1536 ms	& 	4.792 \\
60	& 686.970	& 778.140	& 	779.220	& 	1.133	& 	1.001	& 	5585	ms	& 1684 ms	& 	3.317 \\
65	& 805.520	& 905.660	& 	906.880	& 	1.124	& 	1.001	& 	7349	ms	& 1855 ms	& 	3.962 \\
70	& 930.600	& 1042.970	& 	1044.120	& 	1.121	& 	1.001	& 	5381 ms	& 	2041 ms	& 	2.636 \\
75	& 1054.940& 	1189.970		& 1191.360	& 	1.128	& 	1.001	& 	8480	ms & 	2212	ms	& 3.834 \\
\hline
\end{tabular}
\label{table:5}
\end{table*}

\begin{table*}[t]
\caption{Comparison of Greedy A and Greedy B (N=50, Average over 5 Queries)}
\centering
\begin{tabular}{|c|c|c|c|c|c|} 
\hline
$p$ & $AF_{Greedy A}$  & $AF_{Greedy B}$ \\ 
\hline
3	& 1.030	& 1.000 \\
4	& 1.009	& 1.004 \\
5	& 1.020	& 1.012 \\
6	& 1.059	& 1.018 \\
7	& 1.096	& 1.022 \\
\hline
\end{tabular}
\label{table:6}
\end{table*}

\begin{table*}[t]
\caption{Comparison of Greedy A, Greedy B and LS (Average over 5 Queries)}
\centering
\begin{tabular}{|c|c|c|c|c|c|} 
\hline
$p$ & $AF_{\frac{Greedy B}{Greedy A}}$  & $AF_{\frac{LS}{Greedy B}}$ & $Time_{Greedy A}$ & $Time_{Greedy B}$ & $Time_{\frac{Greedy A}{Greedy B}}$\\ 
\hline
5	& 1.005	& 1	& 1714	& 303	& 5.657 \\
10	& 	1.016	& 	1	& 	1997		& 289	& 	6.910 \\
15	& 	1.036	& 	1	& 	2387		& 381	& 	6.265 \\
20	& 	1.056	& 	1.002	& 	2767		& 522	& 	5.301 \\
25	& 	1.047	& 	1.003	& 	3280		& 574	& 	5.714 \\
30	& 	1.086	& 	1.003	& 	2959		& 537	& 	5.510 \\
35	& 	1.081	& 	1.003	& 	3387		& 622	& 	5.445 \\
40	& 	1.105	& 	1.003	& 	3208		& 704	& 	4.557 \\
45	& 	1.119		& 	1.002	& 	4154		& 837	& 	4.963 \\
50	& 	1.146	& 	1.002	& 	4126		& 1035	& 	3.986 \\
55	& 	1.141	& 	1.002	& 	5559		& 1298	& 	4.283 \\
60	& 	1.156	& 	1.002	& 	5059		& 1411	& 	3.585 \\
65	& 	1.152	& 	1.002	& 	5722		& 1534	& 	3.730 \\
70	& 	1.157	& 	1.002	& 	4766		& 1691	& 	2.818 \\
75	& 	1.151	& 	1.001	& 	7272		& 2180	& 	3.336 \\
\hline
\end{tabular}
\label{table:7}
\end{table*}

\subsection{Experiments on synthetic data sets}
\label{synthetic-greedy} 
Our synthetic data sets are generated by uniformly at random assigning
each vertex $v$ (i.e. element of
the metric space)  a value $f(v) \in [0,1]$, and each distance $d(u,v)$
a value in [1,2]. We note that the \{1,2\} metric is the metric relative to
which the suggested hardness of approximation is derived. We construct
such data sets for various values of $N$, the size of the universe, and for
$p$, the cardinality constraint. In all cases, we set $\lambda = .2$, 
where $\lambda$ is the parameter defining
the relative weight between 
the quality $f(S)$ of a set $S$ and its max-sum dispersion 
$d(S) = \sum_{u,v \in S}d(u,v)$.
For small $N$, we can compute the optimal value and can therefore compute and 
compare the
experimental approximation ratios for Greedy A and Greedy B.


In Table 1 (resp. Table 2), we present results on the relative performance and time 
elapsed for Greedy A, Greedy B, and LS for $N = 50$ (resp. $N = 500$). 
For each setting of the $N,p$ 
parameters we ran 5 trials and averaged the results. We observe these 
average values for each parameter setting for an algorithm $ALG$, and 
report the ``observed average approximation ratio'', 
namely $\frac{OPT-average}{ALG-average}$, denoted $AF_{ALG}$ 
for the $N = 50$ data where we are able to compute the optimum value. 
Similarly, 
we denote the ``relative average approximation'' between two algorithms
as $AF_{\frac{ALG_2}{ALG_1}}$. We also report the average time elapsed
\footnote{The time is reported in milliseconds (ms), with algorithms 
implemented in Java running on a Macbook Pro with 2.4 GHz Intel Core i7 processor and 8 GB 1600 MHz DDR3 memory.} for each algorithm, 
denoted as $T_{ALG}$.  We make the following 
observations based on these trials:

\begin{itemize}
\item Given that max-sum dispersion is a supermodular function, as $N$ grows the
objective value becomes dominated by the dispersion contribution to the diversification result. For each algorithm $ALG$ we show 
its average value $ALG(S)$.  It is observed in our experiments that the max-sum dispersion that is the cause of non-optimality. 

\item
In all cases, the Greedy algorithms and LS perform quite well with regard to
the optimum (when it is computed); this is not surprising as it
is often the case that algorithms perform well for random 
or ``real'' data in contrast to worst case approximation ratios.
More specifically, for $N = 50$ and $p \leq 7$, the approximation ratio for 
GreedyB is roughly $1.02$.  

\item The performance of Greedy A for odd values of $p$ is marred by
the fact, that as defined, Greedy A chooses an arbitrary last vertex rather than
the best last vertex. For larger $p$, this does not have a significant impact 
but it is perhaps best to ignore small odd values of $p$. 
The performance of Greedy 
B is marred by the fact, that as defined, it chooses its first vertex 
arbitrarily rather than choosing a best pair. 

\item As expected, the time bounds for Greedy B are substantially better as
Greedy B is iterating over all vertices rather than over all edges as
in Greedy A.

\item In all cases (for average performance), Greedy B outperforms Greedy A.
For $N = 500$, the relative improvement 
appears generally to be decreasing as $p$ increases, where for the largest 
values of $p = 70$ and $75$, the relative improvement is roughly 
1.5\%. We actually observed in our experiments that the relative improvement was 2.5\% if one 
just compared the dispersion results $d()$.  


\item As expected local search can sometimes improve upon the results of Greedy B, but
obviously at a cost. 
Stopping LS at 10 times the Greedy time, results improve by at most $5\%$ 
and for large $p = 70,75$ the improvement is only $1.5\%$. 

\end{itemize}

Our results for average performance raises the question as to whether or
not Greedy B might outperform Greedy A for all inputs, that is, for all 
parameter settings.  
In order to make the comparison fair, for Greedy A we will 
choose the best final node rather than an arbitrary node when $p$ is odd, and for Greedy B, we
will start with the best pair of nodes rather than an arbitrary node. 
These minor changes do not effect the approximation ratios but can improve the 
observed performance of the algorithms. In Table~\ref{table:3}, we report on these 
improved algorithms, running them for one trial for each of the reported 
parameter setting. We see that 
for $N = 50, p = 4$, there is one setting   
where Greedy A outperformed Greedy B. However, running the algorithms with
these improvements does not alter the basic observations above.  

\begin{table}[t]
\caption{Comparison of documents being returned for the top 50 document data set}
\centering
\subcaption*{N=50, p=3}
\begin{tabular}{|c|c|c|} 
\hline
Greedy A & Greedy B & OPT \\
\hline
4	& 4	& 4 \\
29	& 29	& 29 \\
\textbf{46}	& 24	& 24 \\
\hline
\end{tabular}
\bigskip
\centering
\subcaption*{N=50, p=4}
\begin{tabular}{|c|c|c|} 
\hline
Greedy A & Greedy B & OPT \\
\hline
4	& 4	& 4 \\
29	& 29	& 29 \\
24	& 24	& 24 \\
12	& 12	& 12 \\
\hline
\end{tabular}
\bigskip
\centering
\subcaption*{N=50, p=5}
\begin{tabular}{|c|c|c|} 
\hline
Greedy A & Greedy B & OPT \\
\hline
4	& 4	& 4 \\
29	& 29	& 29 \\
24	& 24	& 24 \\
12	& 12	& 12 \\
\textbf{46}	& 49	& 49 \\
\hline
\end{tabular}
\bigskip
\centering
\subcaption*{N=50, p=6}
\begin{tabular}{|c|c|c|} 
\hline
Greedy A & Greedy B & OPT \\
\hline
4	& 4	& 4 \\
29	& 29	& 29 \\
24	& 24	& 24 \\
12	& 12	& 12 \\
46	& 46	& 46 \\
49	& 49	& \textbf{35} \\
\hline
\end{tabular}
\bigskip
\centering
\subcaption*{N=50, p=7}
\begin{tabular}{|c|c|c|} 
\hline
Greedy A & Greedy B & OPT \\
\hline
4	& 4	& 4 \\
29	& 29	& 29 \\
24	& 24	& 24 \\
12	& 12	& 12 \\
\textbf{0}	& \textbf{49}	& \textbf{37} \\
\textbf{8}	& 46	& 46 \\
\textbf{14}	& 35	& 35 \\
\hline
\end{tabular}
\label{table:8}
\end{table}

\begin{figure}[t]
\centering
    \includegraphics[width=3in]{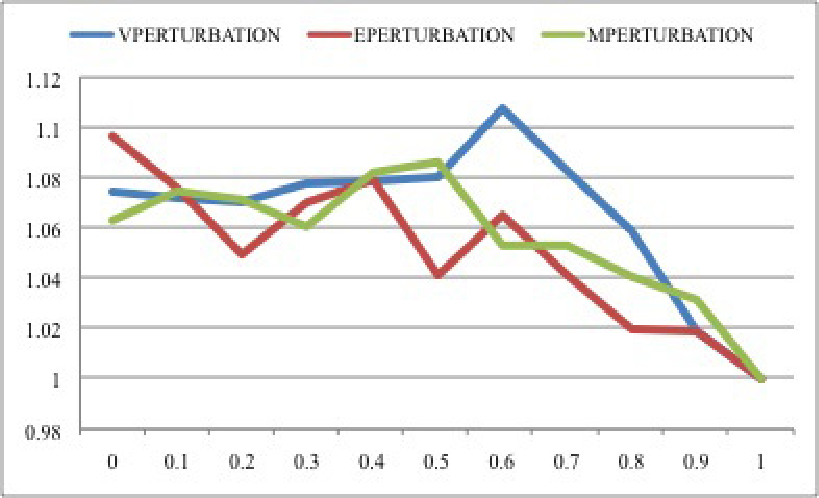}
\caption{Approximation Ratio in Dynamic Updates}
\label{fig:wr}
\end{figure}

\subsection{Experiments with The LETOR data set}
\label{letor-greedy}

We considered popular queries in creating a number of 
 LETOR data sets. 
Each item in a LETOR data set represents a document related to a query. As such,
each item $u$ has an integral relevance score $r(u)$ (relative to the query) 
ranging from 0 to 5, 
a set of feature attributes with their respective values, and a query id. Thus, we take (as ground truth), the quality score $f(S) = \sum_{u \in S} r(u)$. 
We define (and take as ground truth) a metric distance $f(u,v)$ function 
given by the cosine similarity between the feature vectors for $u$ and $v$. 

For Table~\ref{table:4} and Table~\ref{table:5}, we 
chose one data set (chosen at random from the original LETOR dataset) and
created a data set consisting of the top (by relevance score) 
50 and top 370 documents. We applied the Greedy A, Greedy B and limited local search algorithms to these two 
data sets for various settings of the cardinality parameter $p$. 
For the 50 document data set we also computed the optimal values. 
We observe a qualitative 
difference between these ``real data'' experiments and the experiments 
for synthetic data.  
Namely, Greedy B now has a more substantial advantage over 
Greedy A and corresponding decrease in the benefit of running local search for
10 times the run time of Greedy B on a given input. 
In contrast to the synthetic data experiments, the advantage of Greedy B
over Greedy A is more pronounced for larger values of $p$, the cardinality 
constraint. For the $N = 50$ data set and small values of $p$, the advantage 
stays between 3 and 4\%. For the $N = 370$ data set, the advantage of Greedy B 
over Greedy A rises to about 15\% and then levels off at around 12\%. 
The improvement due to the limited use of local search never exceeds .2\% and sometimes results in 
no improvement. We also ran 5 different data sets (i.e. generated by
5 different queries) and averaged the results with respect to the top 50 results and all documents in each data set
as shown in Table~\ref{table:6} and Table~\ref{table:7}. Note that in these tables, we are omitting the objective function 
values that have been previously included in other tables. We are averaging our results over different LETOR datasets (i.e. queries) and therefore reporting on the average objective function values wont be fully meaningful. These average results support what we found in Table~\ref{table:4} and Table~\ref{table:5}, namely that Greedy B significantly outperforms Greedy A and that limited local search provides a  very small advantage over Greedy B. 

In Table~\ref{table:8}, we present the difference in the documents being returned for
the 50 document data set. Here the OPT documents are the true
set of optimal documents with respect to the diversification function applied
to the values of  the document relevance scores and the cosine 
distance function. As an example, consider the results for the $N = 50, p = 7$
setting of the parameters. Here OPT and Greedy B differ on one document
while Greedy A differs on 3 documents.  

\subsection{Approximation Ratio in Dynamic Updates}
\label{dynamic-experiments}
For dynamic updates, we use same synthetic data as in 
Section~\ref{synthetic-greedy}. 
We have three different dynamically changing environments:
\be
\item {\sc vperturbation}: each perturbation is a weight change on an item;
that is, an item (vertex) $u$ is randomly chosen and its value 
value is reset uniformly 
at random   
from $[0,1]$.
\item {\sc eperturbation}: each perturbation is a distance change between two 
items; that is, a pair of distinct items $\{u,v\}$ is randmoy chosen 
and the distance $d(u,v)$ is reset uniformaly at random from $[1,2]$.  
\item {\sc mperturbation}: each perturbation is one of the above two with equal probability.
\ee
For each of the environments above and every value of $\lambda$, we start with our  greedy solution (a 2-approximation) and run 20 steps of simulation, where each step consists of
a random weight change of the stated type, followed by a single application of the oblivious update rule.
We repeat this 100 times and record the worst approximation ratio occurring during these 100 updates. The results are shown in Fig.~\ref{fig:wr}; the horizontal axis measures $\lambda$ values, and the vertical axis measures the approximation ratio.

We have the following observations:
\be
\item In any dynamic changing environment, the maintained ratio is well below the provable ratio of 3.
The worst observed ratio is about 1.11.
\item The maintained ratios are decreasing to 1 for increasing
$\lambda \geq 0.6$. 
\ee
From the experiment, we see that the simple local search update rule seems effective for maintaining a good approximation ratio in a dynamically changing environment.

%% file: conclusion-rev.tex
\section{Conclusion}
We study the max-sum diversification with monotone submodular set functions and give a natural 2-approximation
greedy algorithm for the problem when there is a cardinality constraint. 
We further extend the problem to matroid constraints and give a 2-approximation
local search algorithm for the problem.  
We examine the dynamic update setting for modular set functions, where the 
weights and distances are constantly changing over time and the goal is to maintain a solution with good quality with a limited number of updates. We propose a simple update rule: the oblivious (single swap) update rule, and show that if the weight-perturbation is not too large, we can maintain an approximation ratio of 3 with a single update. The diversification problem has many important applications and there are many interesting future directions. 
Although in this paper we restricted ourselves to the max-sum objective, there are many other well-defined notion of diversity  that can be considered, see for example~\cite{Chandra:1996:FDR:645898.756652} and \cite{GoSh09}.  The max-sum case can be also studied for specific metrics such as the $\ell_1$-norm 
in Euclidean space as 
considered by Fekete and Meijer \cite{FeketeM03} who provide a linear
time optimal algorithm for constant $p$ and a PTAS when $p$ is
part of the input. Their PTAS 
algorithm also provides a $(2+\epsilon)$-approximation for the $\ell_2$-norm.
Their algorithms exploit the geometric nature of the metric space. Other 
specific metric spaces are also of interest. 

In the general matroid case, the greedy algorithm given in Section~\ref{sec:submo} fails to achieve any constant approximation ratio, but one can also 
consider other ``greedy-like algorithms'' such as the partial enumeration
greedy method used (for example) successfully for monotone submodular 
maximization subject to
a knapsack constraint in Sviridenko \cite{Sviridenko04}? Can such a technique
also be used to provide an approximation for our diversification problem? 
Can our results be extended to provide a constant approximation for
the diversification problem subject to a knapsack constraint? 
In a dynamic update setting, we only considered the oblivious 
single swap update rule. 
It is interesting to see if it is possible to maintain a better ratio than 3 with a limited number of updates, by larger cardinality swaps, and/or by
a non-oblivious update rule. We leave this as an open question.
The approximation ratio and application of diversification maximization in a 
distributed setting is pursued in a recent paper by Abbasi-Zadeh et al 
\cite{AZadehGMZ16}.

Finally, a crucial property used throughout our results 
is the triangle inequality. 
In our conference paper \cite{BorodinLY12}, we asked the question as to
whether we 
can relate the approximation ratio to the parameter of a relaxed triangle inequality? Sydow \cite{Sydow14} provides a partial answer to this question 
showing that the matching based algorithm of Hassin et al \cite{HassinRT97}  
can be applied to an $\alpha \geq 1$ relaxed metric (where 
$d(x,y) + d(y,z) \geq \alpha d(x,z))$ resulting in a (tight) 
$\frac{2}{\alpha}$ approximation ratio for the cardinality constrained 
max-sum dispersion problem. More recently and independently, Abbasi-Zadeh and 
Ghadiri \cite{AZadehG15} obtain the $\frac{2}{\alpha}$ approximation ratio 
for the cardinality constraint, and a $\frac{2}{\alpha^2}$ approximation ratio 
for an arbitrary matroid constraint.

%% file: appendix.tex
\section{The Greedy Algorithm Applied to Diversification with a Matroid 
Constraint}
\label{greedy-matroid}
We observe that 
for the more general matroid constraint diversification problem, 
the greedy algorithm in section \ref{sec:submo} no longer achieves any constant approximation ratio. More specifically, consider the max-sum diversifciation problem
as in Gallopudi and Sharma \cite{GoSh09} (that is, for a modular quality
function $f()$)  but now subject to a partition matroid constraint.
Partition the universe into
$A = \{a,b\}$ with cardinality constraint 1 and $C = \{c_1, c_2, \ldots c_r\}$ with no cardinality
constraint. Let the objective be
$f(S) = \sum_{u \in S} q_u + \sum_{u,v \in S} d(u,v)$
where the quality and distance functions are defined as follows:
$q(a) = \ell + \epsilon$,  $q(x) = 0$ for all $x \neq a$, and for all $x$,
$d(b,x) = \ell$, $d(u,x) = \epsilon $ for all $u \neq b$. The greedy algorithm
(starting with $a$ or with the best pair $(a,b)$ will yield
$f(S) = f(C \cup \{a\}) =  \ell + \epsilon + \epsilon \cdot {r \choose 2} + r \epsilon$
while the optimal solution
will be $f(C \cup \{b\}) =  r \cdot \ell + \epsilon \cdot {r \choose 2}$. Hence the
approximation can be made arbitrarily bad by choosing
$\epsilon = \frac{1}{{r \choose 2}}$ and making $r$ sufficiently large.

By the reduction in \cite{GoSh09} to the metric dispersion problem, the above example
shows that the greedy algorithm will also suffer the same unbounded
approximation ratio for the metric dispersion problem. 

